\documentclass[aps,prd,showpacs,preprintnumbers,nofootinbib,twocolumn]{revtex4}

\usepackage{bm}
\usepackage{latexsym}
\usepackage{dcolumn}
\usepackage{amsfonts,amssymb}
\usepackage{graphicx,epsfig}
\usepackage{psfrag}
\usepackage{amsmath,amssymb}

\def\beq{\begin{equation}}
\def\eeq{\end{equation}}
\def\br{\begin{eqnarray}}
\def\er{\end{eqnarray}}
\def\benu{\begin{enumerate}}
\def\eenu{\end{enumerate}}
\def\nn{\nonumber} 
\def\pa{{\partial}}
\def\l{\left}
\def\r{\right}

\def\lPl{\ell_{_{\rm Pl}}}

\def\a  {\alpha}
\def\b  {\beta}
\def\c  {\gamma}

\def\e  {\epsilon}

\def\k  {\kappa}

\def\m  {\mu}
\def\n  {\nu}
\def\o  {\omega}
\def\O  {\Omega}
\def\p  {\pi}

\def\s {\sigma}

\def\la {\label}
\def\pa {\partial}
\def\ba{\begin{eqnarray}}
\def\ea{\end{eqnarray}}
\def\f {\frac}

\def\bi {\begin{itemize}}
\def\ei {\end{itemize}}

\def\le {\left}
\def\ri {\right}
\def\be {\begin{equation}}
\def\ee {\end{equation}}
\def\la {\label}
\begin{document}

\preprint{hep-th/0410209}

\title{High frequency quasi-normal modes for black-holes with generic 
singularities}
\author{Saurya Das}
\email[]{E-mail: saurya.das@uleth.ca}
\affiliation{Department of Physics, University of Lethbridge,\\
4401 University Drive, Lethbridge, Alberta T1K 3M4, CANADA}
\author{S.~Shankaranarayanan}
\email[]{E-mail: shanki@ictp.trieste.it}
\affiliation{HEP Group, The Abdus Salam International Centre 
for Theoretical Physics,\\ Strada costiera 11, 34100 Trieste, Italy.}
\date{\today}


\begin{abstract}
We compute the high frequency quasi-normal modes (QNM) for scalar
perturbations of spherically symmetric single horizon black-holes in
$(D+2)$-space-time dimensions with generic curvature singularities and
having metrics of the form $ds^2 = \eta x^p (dy^2-dx^2) + x^q d\O_D^2$
near the singularity $x=0$. The real part of the QN frequencies is
shown to be proportional to $\log \le[ 1 + 2\cos \le( \p \le[ qD -2
\ri]/2 \ri ) \ri]$ where 
the constant of proportionality is equal to the 
Hawking temperature for non-degenerate black-holes and inverse of
horizon radius for degenerate black-holes. Apart from agreeing with
the QN frequencies that have been computed earlier, our results imply
that the horizon area spectrum for the general spherically symmetric
black-holes is equispaced. Applying our results, we also find the QNM
frequencies for extremal Reissner-Nordstr\"om and various stringy
black-holes.
\end{abstract}
\maketitle


\section{Introduction} 

Quasi-normal modes (QNM) are classical perturbations with non-vanishing
damping propagating in a given gravitational background subject to
specific boundary conditions. The frequency and damping of these
oscillations depend only on the parameters characterizing the black
hole and are completely independent of the particular initial
configuration that caused the excitation of such vibrations. Over the
last three decades, QNM have been of interest due to their
observational significance in the detection of gravitational
waves. (For a review, see Ref. \cite{Kokkotas-Schm:1999})

During the last few years there has been renewed interest in QNM for
the following two reasons. First, in estimating the thermalization
time-scales in connection with the AdS/CFT conjecture
\cite{Horowitz-Hube:1999}. Secondly, and more importantly, it has been
conjectured recently that the real part of the QN frequencies for
black-hole perturbations ($\o_{QNM}$) correspond to the minimum energy
change of a black-hole undergoing quantum transitions
\cite{Hod:1998,Dreyer:2002}.

Earlier, Bekenstein \cite{Bekenstein:1974,Bekenstein-Mukh:1995} had
conjectured using semi-classical arguments that the black-hole area
spectrum is equispaced and is of the form
\begin{equation}
A_s = 4\, \log(k)\, \lPl^2 \, s \qquad s = 1,2, \cdots \, , 
\label{eq:arean}
\end{equation}
where $k$ is an integer to be determined and $\lPl$ is the Planck
length. This implies that for a $(D + 2)-$dimensional Schwarzschild
black-hole, the energy $\omega = \Delta M$ emitted when the black-hole
looses one quantum of the area is
\beq
\omega = \f{(D - 1)\ln k}{4\p r_h} = \le({\ln k}\ri){T_H} \, ,
\label{eq:freq}
\eeq
where $r_h$ is the horizon and $T_H$ is the Hawking temperature. 

Hod \cite{Hod:1998} noticed, based on the numerical result
\cite{Nollert:1993}, that the QNM spectrum for the Schwarzschild
black-hole has a frequency whose real part numerically approached
Eq. (\ref{eq:freq}) with $k = 3$ in the limit of infinite imaginary
frequency. This was later confirmed analytically by Motl
\cite{Motl:2002} for $4$-dimensional
Schwarzschild, who showed that the QN frequencies have the following
structure:
\beq
\omega_{_{QNM}} = 2 \pi \, i\, T_H \l(n + \frac{1}{2}\r)  + T_H \ln(3) 
+ {\cal O}(n^{-1/2}) \, .
\la{mn01}
\eeq

Eqs.~(\ref{eq:arean}) and (\ref{mn01}) were used by Dreyer
\cite{Dreyer:2002} recently to argue that the Barbero-Immirzi
parameter appearing in loop quantum gravity is $\ln(3)/(2\sqrt{2}\p)$
and that the spin network states in the theory are dominated by
$j_m=1$ representations of the $SU(2)$ group\footnote{Note, however,
that the above value of the Barbero-Immirizi parameter does not agree
with that predicted by the recent analyses by Domagala and Lewandowski 
\cite{Domagala-Lewa:2004}, and Meissner \cite{Meissner:2004}. 
See also Refs. \cite{Dreyer-Mark:2004,Alexandrov:2004}}. Together,
this also led to the microscopic entropy of the black-hole to be one
quarter its horizon area (in Planck units). Turning the argument
around, (\ref{eq:arean}) and (\ref{mn01}) can also be used to
vindicate the point of view that horizon area is equispaced
\cite{Bekenstein-Mukh:1995,Kunstatter:2002}.

Subsequently, Motl and Neitzke \cite{Motl-Neit:2003} (see also
Ref.~\cite{Andersson-Howl:2003}) used a more flexible and powerful
approach, called the monodromy method, and were able to compute
high-frequency QNM and the asymptotic value of the real part of the QN
frequencies for a $D$-dimensional Schwarzschild.  Naturally, attempts
were made to compute the QN frequencies for a host of other
black-holes.  There have been quite a bit of effort to understand the
physics underlying the real (see, for instance,
Refs. \cite{Suneeta:2003,Medved-Martin:2003,Paddy:2003,Tirth-Paddy:2003,Cardoso-Phd:2003,Kettner-Kuns:2004})
and the imaginary part of the high-frequency QNM (see, for instance,
Refs.
\cite{Govindarajan-Sune:2000,Birmingham:2003,Birmingham-Carp:2003,Birmingham-Sach:2001,Berti-Cardoso:2004,MaassenvandenBrink:2003,Musiri-Siopsis:2003c,Oppenheim:2003,Cardoso-Kono:2003,Cardoso-Lemo:2003,Konoplya:2002,Konoplya:2003,Setare:2003,Konoplya:2004,Setare:2004}),
and there appears to be pieces of evidence in favour of as
well as against the predictions referred to above. Thus, it is
important to explore the QN frequencies for other black-holes and to
probe their full implications.

Even though the monodromy approach does not require the full knowledge
of the space-time -- except at the singularity, horizons and spatial
asymptotic infinity -- all the previous analyses have been restricted 
to space-times whose 
line-element is known for the whole of the manifold including the
recent work by Tamaki and Nomura \cite{Tamaki-Nomu:2004}. In this
work, we compute the QN frequencies for $(D+2)-$dimensional
spherically symmetric single horizon black-holes with generic
singularities and near-horizon properties (which include the ones that
have been already explored). Near the horizon, we assume that the
spherically symmetric metric takes the form of the Rindler while close
to the singularity we use the form of Szekeres-Iyer metric
\cite{Szekeres-Iyer:1993,Celerier-Szek:2002,Blau-Boru:2004}.
Using the Monodromy approach, we show that (i) the imaginary part of
the high frequency QNM are discrete and uniformly spaced and (ii) the
real part depends on the horizon radius, space-time dimension and
power-law index of $S^D$ near the singularity. We also show that the
real part of the high frequency QNM has a logarithmic dependence whose
argument need not necessarily be an integer. In order to illustrate this
fact, we consider specific black-holes and obtain their QN
frequencies.

The rest of the paper is organized as follows. In the next section, we
discuss generic properties of the space-time near the horizon and the
singularity. In Sec. (\ref{sec:qnmcalc}), we obtain the 
QN frequencies for a general static spherically symmetric black-holes.
In Sec. (\ref{sec:specificBH}), we apply our general results to
specific black-holes, reproducing earlier results and obtaining new
ones. Finally, we conclude in Sec. (\ref{sec:conc}) summarising our
results and speculating on future directions.

\section{Spherically symmetric black-hole}
\label{sec:SphSy}

We start with the general $(D+2)-$dimensional spherically symmetric
line element:
\br
\label{eq:gen-4D}
ds^2
\label{eq:spher-tr}
&=& - f(r) dt^2 + \frac{dr^2}{g(r)} + \rho^2(r) d\Omega_D^2 \, ,\\ 
\label{eq:spher-tx}
&=&  f(r)\l[ -dt^2 + dx^2\r] + \rho^2(r) d\Omega_D^2 \, ,
\er
where $f(r)$, $g(r)$ and $\rho(r)$ are arbitrary (continuous,
differentiable) functions of the radial coordinate $r$, $d\Omega_D^2$
is the metric on unit $S^D$ and 
\beq
x = \int \frac{dr }{\sqrt{f(r) g(r)}} ~,
\label{eq:rela-xr}
\eeq
denotes the tortoise coordinate. As it can be seen, the line-element
(\ref{eq:spher-tx}) factorizes into the product of two spaces ${\cal
M}^2 \times S^D$, where ${\cal M}^2$ is the $2$-dimensional
space-time with Minkowskian topology.

In order for the line-element (\ref{eq:spher-tr}) to describe a static
black-hole, the space-time must contain a singularity (say at $r = 0$)
and have horizons. In this work, we will assume that the space-time
contains one event-horizon at $r_h$ and that it is asymptotically
flat. However, we do not assume any specific form of $f(r)$, $g(r)$
and $\rho(r)$.  In the rest of this section, we discuss generic
properties of the space-time near the horizon $(r = r_h)$ and the
singularity $(r = 0)$.

\subsection{Horizon structure}

For the line-element (\ref{eq:spher-tr}), there exists a time-like
Killing vector field $\xi^{\mu}(x^{\nu})$ which can be expressed as
$\xi^{\mu}(x^{\nu}) = (1, 0,\cdots,
0)$\cite{Wald:1984-bk}. Substituting $I^{\mu} = \xi^{\mu}$ into the
definition of the surface gravity, i. e.,
\beq
\kappa^2 = - \frac{1}{2} I^{\mu,\nu} I_{\mu,\nu} 
\eeq
we obtain \cite{Wald:1984-bk}
\beq
\kappa = \frac{1}{2} \le(\sqrt{\frac{g(r)}{f(r)}} 
\frac{df(r)}{dr}\ri)_{r = r_h} \, .
\label{eq:kappa-def}
\eeq
Using the property that the event horizon is a null hypersurface, the
location of the horizon is determined by the condition $g^{\mu\nu} \,
\pa_{\mu}N \, \pa_{\nu}N = 0$.  For the line-element
(\ref{eq:spher-tr}) $N$ is a function of $r$ characterizing the null
hypersurface which gives $g(r_h) = 0$.  Thus, the location of the
event horizon is given by the roots of the above equation.

Since $\kappa = constant$ and $g(r) = 0 $ at the event horizon, using 
the definition of surface gravity from Eq. (\ref{eq:kappa-def}), we have 
the condition $f(r_h)/g(r_h) = H(r_h)$ where $H(r_h) \neq 0$. Using the 
property that $f(r)$ and $g(r)$ are smooth functions, we have the following 
relation for general black-holes:
\beq
f(r)/g(r) = H(r)
\eeq
where $H(r)$ is a smooth function and is non-vanishing at the
event-horizon.

In order to obtain the line-element near the horizon, we make the 
coordinate transformation $(t, r)~\to~(t,\gamma)$, which is 
defined by
\beq
\gamma \equiv \frac{1}{\kappa} \sqrt{f},\qquad d\gamma = 
\frac{1}{2\kappa} \frac{d_r f}{\sqrt{f}} \, dr \, ,
\eeq
where $\kappa$ is given by (\ref{eq:kappa-def}). Note that the
horizon ($r_h$) is at $\gamma = 0$. The line-element
(\ref{eq:spher-tr}) becomes
\beq
ds^2 = - \kappa^2 \gamma^2 dt^2 + 4 \, \frac{f}{g} \frac{\kappa^2}{(d_r f)^2}
d\gamma^2 + \rho^2(r) d\Omega^2_D \, ,
\eeq
and hence, near the horizon, we have
\beq
ds^2 \to - \kappa^2 \gamma^2 dt^2 + d\gamma^2 + \rho^2(r_0) \, d\Omega^2_D \, .
\eeq
For space-times with single non-degenerate horizon (like Schwarzschild
for which $\kappa \neq 0$), $f(r), g(r)$ can be expanded around $r_h$ as
\beq
f(r) = f'(r_h) ~ (r - r_h); ~g(r) = g'(r_h) ~ (r - r_h) \, ,
\label{eq:fg-rh}
\eeq
using (\ref{eq:kappa-def}), we have
\beq
\k =  \frac{1}{2} \sqrt{g'(r_h) \, f'(r_h)} \, ,
\eeq
and using the relation (\ref{eq:rela-xr}), we have
\beq
x = c_0 \, \ln(r - r_h) \, ,
\label{eq:rel-xrh}
\eeq
where $c_0$ is a constant and is given in Table (\ref{ta:1}).

In the case of $(D + 2)-$dimensional Schwarzschild, we have 
\beq
f(r) = g(r) = 1 - \left(\frac{r_h}{r}\right)^{D - 1},
\eeq
where $r_h$ is related to the black-hole mass $(M)$ by the relation $M
= (D \, \Omega_D \, r_h^{D - 1})/(16 \pi G_{D+2})$,
$G_{D+2}$ being the $(D+2)-$dimensional Newton's constant and 
$\Omega_D = (2 \pi^{(D + 1)/2}/\Gamma[(D + 1)/2]$. Using 
(\ref{eq:kappa-def}), we have
\beq
\kappa = \frac{(D - 1)}{2 \, r_h} \, .
\eeq
For space-times with degenerate horizon, such as the extremal
Reissner-Nordstr\"om (RN), we have 
\beq
f(r) = g(r) = \frac{(r^{D - 1} - r_h^{D - 1})^2}{r^{2(D - 1)}}
\label{eq:RN-fr}
\eeq
where 
\beq
r_h = \l(8\p G_{D+2} M\r)/\l(D~\O_D\r) 
\eeq
For these space-times using the relation (\ref{eq:rela-xr}), we have
(near the horizon)
\beq
x \simeq c_0 \log(r - r_h) + {\mathcal O}(r^n)
\eeq
where $c_0$ is a constant and is given in Table (\ref{ta:1}). See
Table (\ref{ta:1}) and also Sec. (\ref{sec:specificBH}) for properties
of the stringy black-holes.

\subsection{Generic power-law singularities}

The analysis of Motl-Neitzke \cite{Motl-Neit:2003} depends crucially
on the behavior of the metric near the singularity. Thus, in order to
assess the generality of the result, one needs to understand the
generality of the space-time singularities. A decade ago,
Szekeres-Iyer
\cite{Szekeres-Iyer:1993,Celerier-Szek:2002} 
(see also, Ref. \cite{Blau-Boru:2004}) investigated a large class of
four-dimensional spherically symmetric space-times with power-law
singularities. These space-times practically encompass all
known spherically symmetric solutions of the Einstein equations such
as Schwarzschild(-de Sitter), Reissner-Norstr\"om and other type of
metrics with null singularities. In Sec. (\ref{sec:specificBH}), we
will show that certain stringy black-holes and higher dimensional
Gibbons-Maeda type black-holes
\cite{Gibbons-Maed:1987,Garfinkle-Horo:1991,Horowitz-Mald:1996} also
fall into this class.

Szekeres-Iyer had shown that near the singularity, the spherically
symmetric metric (\ref{eq:spher-tr}) takes the following form(s):
\br
\label{eq:singl-line}
\!\!\!\!\! ds^2 &\stackrel{r \to 0}{\simeq}& \eta r^{2p/q} dy^2 - \frac{4\eta}{q^2}r^{2(p-q+2)/q}dr^2 + r^2
d\Omega_D^2 \, , \\
\label{eq:singl-line2}
&=& \eta x^p (dy^2 - dx^2) + x^q d\O_D^2 \, ,
\er
where $y=\b t,~\b>0$, $\eta = 1, 0, -1$ correspond to space-like, null
and time-like singularities respectively and $p, q$ are constants and
capture the dominant behavior near the singularity. Note that the
notation of $t$ and $r$ is adapted to the case of $\eta = -1$ where
the singularity is time-like and $t$ is time. However, we will
continue to use this notation even for space-like singularities where
$t$ is actually space-like. The line-element (\ref{eq:singl-line2})
clearly shows that near the singularity the product spaces have
different singularity structure. 
The curvature invariants -- Ricci 
and Kretschmann scalars -- for the line-element (\ref{eq:singl-line})
go as
\br
\la{curv1}
R &=& \f{ax^{-p}+bx^{2-q}}{x^2}~, \\
R_{abcd}R^{abcd} &=& \f{c x^{-2p} + d x^{-2q+4} + e x^{-p-q +2}}{x^4},
\la{curv2}
\er
where $a(p,q),\dots,e(p,q) = {\cal O}(1)$. The form of the invariants
show that the Szekers-Iyer line-element indeed describe the
spherically symmetric space-time near the singularity.

Comparing Eqs. (\ref{eq:spher-tr}, \ref{eq:singl-line}), we have 
\beq
\label{eq:fg-r0}
\!\!\! f(r) = - \eta \b^2 r^{2p/q};~ 
\frac{1}{g(r)} = - \frac{4\eta}{q^2}r^{2(p-q+2)/q};~
\rho(r) = r \ .
\eeq
Substituting the above expressions in Eq. (\ref{eq:rela-xr}), we have 
\beq
x = \f{r^{2/q}}{\b} \, . 
\label{eq:rel-xr0}
\eeq
Note that near the curvature singularity, the tortoise coordinate
depends on $q$ but not on $p$. This will be crucial in obtaining the
real part of the high frequency QNM. In Table (\ref{ta:1}), we have
given the values of $p, q$ for various black-holes.

For example, in the case of $(D+2)$-dimensional Schwarzschild:
\be
p = \frac{1 - D}{D}~;\quad  q = \frac{2}{D}~; \quad x = \f{r^D}{\b}~,
\ee
whereas for $(D+2)$-dimensional (non-)extremal Reissner-Nordstr\"om:
\be
p = - 2 \le( \f{D-1}{2D-1} \ri)~;\quad q = \f{2}{2D-1}~; \quad 
x = \f{r^{2D-1}}{\b}~.
\ee

\section{Quasi-normal modes for static black-holes}
\label{sec:qnmcalc}

In this section, we obtain the high (imaginary) frequency QNM,
corresponding to the scalar gravitational perturbations, for the
general spherically symmetric $(D+2)-$dimensional black-holes
discussed in the previous section.

\subsection{Scalar Perturbations}

The perturbations of a $(D + 2)$-dimensional static black-holes
(\ref{eq:spher-tr}) can result in three kinds -- scalar, vector and
tensor -- of gravitational perturbations (see for example,
Ref. \cite{Kodama-Ishi:2003}). The higher dimensional scalar gravitational 
effective potential, which is of our interest in this work, correspond
to the well-known four-dimensional Regge-Wheeler potential. The
evolution equation for the scalar gravitational perturbations follows
directly from the massless, minimally coupled scalar field propagating
in the line element (\ref{eq:spher-tr}), i. e.,
\br
& & \Box~\Phi \equiv \frac{1}{\sqrt{-g}}~
\pa_{\mu} \le( \sqrt{-g}g^{\mu\nu}\pa_{\nu} \Phi \ri) =  0 \, ,\nn \\
& & 
\frac{d^2 R(r)}{dx^2} + \le[ \omega^2 - V(r) \ri] R(r) = 0 \, ,
\label{eq:wav-gen-spher}
\er
where $\Phi(x^{\mu}) = \rho(r)^{-D/2} \, R(r) \, \exp(i \omega t) \, Y_{l
m_1 ... m_{D - 1}}$ and $V(r)$ is the higher dimensional
analog of the Regge-Wheeler potential and is given by
\br
V(r) &= & \f{l(l+D-1)}{r^2}~f(r) 
+ \le( \f{D}{2} \ri) \rho(r)^{-\f{D}{2}} \sqrt{f(r) \,g(r)}~ \nn \\
\label{eq:RW-gen}
&\times& \f{d}{dr} \le( \rho(r)^{\f{D-2}{2}} \, \frac{d\rho(r)}{dr} \sqrt{f(r) \, g(r)} \ri) \, .
\er
The analysis of Motl and Neitzke requires the extension of 
Eq. (\ref{eq:wav-gen-spher}) beyond the physical region $r_h < r <
\infty$. In order to perform the analysis we need to know the 
nature of the singularity of the differential equation
(\ref{eq:wav-gen-spher}) at $r = 0, r_h,~{\rm and}~ \infty$.  Assuming
that the space-time is asymptotically flat at radial infinity and
using the relation (\ref{eq:fg-rh}) near the horizon, it is easy to
shown that $r \to \infty$ and $r = r_h$ are irregular and regular
singular points of the differential equation (\ref{eq:wav-gen-spher})
respectively. In the case of power-law singularities as discussed in
the previous section, in order for $r = 0$ to be a regular singular
point of the differential equation (\ref{eq:wav-gen-spher}), it can be
easily shown that $p, q$ {\it must} satisfy the following conditions:
\beq
q>0 ~~\mbox{and}~~p - q + 2 > 0  
\label{eq:pq-cond}
\eeq
The above conditions will be useful in reducing the generalized
Regge-Wheeler potential (\ref{eq:RW-gen}) near the singularity similar
to the near-origin form of the potential derived by Motl and Neitzke
\cite{Motl-Neit:2003}. Using the above conditions and $x$ continued to
the whole complex plane (say $z$), Eq. (\ref{eq:wav-gen-spher}) is an
ordinary differential equation with regular singular points at $r =
0,~ r_h$ and an irregular singular point at $r = \infty$. Thus, by the
general theory of differential equations \cite{Morse-Fesh:1953-bk1},
any solution of (\ref{eq:wav-gen-spher}) in the physical region
extends to a solution on the $r-$plane. However, this solution may be
multi-valued around the singular points at $r = 0,~r_h$
\cite{Motl-Neit:2003}. 

\subsection{Computation of quasi-normal modes}

QNM are solutions to the differential equation (\ref{eq:wav-gen-spher}) whose
frequency is allowed to be complex. In the case of asymptotically flat
space-times, which is of our interest, the modes are required to have
purely outgoing boundary conditions both at the horizon and in the
asymptotic region, i. e.,
%
\beq
R(x)\sim e^{\pm i\omega x}\quad {\rm as} \quad x\rightarrow \mp \infty\, .
\label{eq:bc}
\eeq
In order for the black-hole to be stable, the modes should decay in
time, hence $\Im(\omega) > 0$. In the monodromy approach
\cite{Motl-Neit:2003}, unlike the earlier approaches, the authors
analytically continued $x$ (in the complex plane $z$), instead of
$\omega$, and introduced the boundary conditions as the product
$\omega \, z\rightarrow\pm\infty$, instead of $x \rightarrow
\pm\infty$. In this approach, we need to know the solution of
Eq. (\ref{eq:wav-gen-spher}) near $r = 0,~r_h$ and compare their
monodromies.

For the general spherically symmetric space-time -- with the power-law
singularity at the origin and generic horizon structure -- the
generalized Regge-Wheeler potential (\ref{eq:RW-gen}) near $r = 0$ and
$r = r_h$ is
\begin{widetext}
\br
\label{eq:RW-rh}
V(r) &\stackrel{r \to r_h}{\simeq}& 
\l[ \frac{l (l + D - 1)}{\rho^2(r_h)} \, f'(r_h)
+ \frac{D}{2}\,{f'(r_h)\, g'(r_h)} \, \frac{\rho'(r_h)}{\rho(r_h)}\r]\, (r - r_h) 
+ {\cal O}[(r - r_h)^2] \, , \\
\label{eq:RW-r0}
%
&\stackrel{r \to 0}{\simeq}& 
\le( \f{q\b}{2} \ri)^2~\f{D}{2}~\le( \f{D}{2} - \f{2}{q} \ri)~r^{-4/q}
- \eta \b^2 \, l \,(l + D - 1) \, r^{2 (p - q)/q} \, . 
\er
\end{widetext}
Following points are worth noting regarding the above result:\\ (i)
Using the conditions (\ref{eq:pq-cond}) in the second equation above,
it is clear that near the origin the first term in the RHS dominates
the second term. Hence, it would suffice to consider first term in the
rest of the analysis. Rewriting the potential near the origin in-terms
of $z$, we have
\beq
V[r[z]] \stackrel{r \to 0}{=} \frac{D \, q}{8} \l( \frac{q D}{2} - 2 \r) 
\frac{1}{z^2} \, .
\label{eq:V-rz0}
\eeq
(ii) It is also interesting to note that the generalized Regge-Wheeler
potential near the origin depends {\it only} on $q$, $D$. From the
line-element (\ref{eq:singl-line2}), this implies that the potential
depends on the singularity structure of the $S^{D}$ space and not on
${\cal M}^2$. \\ (iii) Remarkably, the generalized Regge-Wheeler
potential near the origin is similar to the form of the near-origin
potential derived by Motl and Neitzke \cite{Motl-Neit:2003} except for
the coefficients\footnote{The above behaviour of the Regge-Wheeler
potential was also observed in Ref. \cite{Blau-OLou:2004}. Other
consequences of this universal behaviour will be explored in
Ref. \cite{Blau-Fran:2004}.}. In view of these observations, the
monodromy calculation
\cite{Motl-Neit:2003} should carry through relatively unchanged 
for the general spherically symmetric metrics with power-law
singularities. In the rest of the section, we describe this
calculation for the general spherically symmetric metrics and obtain
the high (imaginary) frequency QNM. [We follow the notation of Motl
and Neitzke closely to provide easy comparison.]

Substituting the generalized Regge-Wheeler potential (\ref{eq:V-rz0})
in the differential equation (\ref{eq:wav-gen-spher}), we get 
\beq
R(z)=A_{+}c_{+}\sqrt{\omega z}J_{+\nu}(\omega z)
+A_{-}c_{-}\sqrt{\omega z}J_{-\nu}(\omega z)
\;, 
\label{eq:solr=0}
\eeq
where $\nu = (D\,q - 2)/4$, while the products $c_{+}A_{+}$ and
$c_{-}A_{-}$ represent constant coefficients. Following
\cite{Motl-Neit:2003}, we will choose the ``normalization factors'' (denoted
by $c_{\pm}$) so that
\beq
c_{\pm}\sqrt{\omega z}J_{\pm\nu}(\omega z)
\sim 2\cos\left(\omega z-\alpha_{\pm}\right)
\quad {\rm as} \quad\omega z\rightarrow\infty\;,
\label{eq:cpm}
\eeq
with
\beq
\alpha_{\pm}\equiv \frac{\pi}{4}\left[1\pm 2\nu\right] \, .
\label{eq:alpha}
\eeq
{}From Eqs. (\ref{eq:solr=0}) and (\ref{eq:cpm}), as well as the
boundary condition (\ref{eq:bc}) in-terms of $\omega\,z$  we
get the following constraint
\beq
A_+e^{-i\alpha_{+}}+A_-e^{-i\alpha_{-}}=0
\label{eq:constApm}
\eeq
and obtain the asymptotic form for $R(z)$ as 
\beq
R(z)\sim \left[A_+e^{+i\alpha_{+}}+A_-e^{+i\alpha_{-}}\right]e^{-i\omega z}
\quad {\rm as}\quad \omega z\rightarrow \infty \;.
\label{eq:asymptR1}
\eeq
We follow the contour from the the negative imaginary axis of $z$
($\Im (\omega z)\rightarrow\infty$) to the positive imaginary axis
($\Im (\omega z)\rightarrow -\infty$). 
Now from 
Eq.(\ref{eq:rel-xr0}) and the relation $z=0$, it follows that the 
the rotation of the contour on the $r$ plane is by an angle 
$3\p q/2$, i.e. $r \rightarrow r \exp({i3\p q/2})$.
This translates to a rotation of $3\p q/2 \times 2/q
= 3\p$ in the $z$ plane. That is: $z \rightarrow \exp(3i\p) z$.  
%
Further, using $J_\n(z e^{im\p}) = e^{im\n\p}J_\n(z)$, we
have (as $\omega z\rightarrow -\infty$)
\br
R(z) &\sim& \left(A_+ e^{5 i\alpha_{+}}
+ A_- e^{5 i\alpha_{-}}\right)e^{-i\omega z}
\label{eq:asymptR2}
\\ 
&+&\left(A_+ e^{7 i\alpha_{+}}
+ A_- e^{7 i\alpha_{-}}\right)e^{+i\omega z}
\nn \;.
\er
Then, from Eqs. (\ref{eq:asymptR1},\ref{eq:asymptR2}), following 
\cite{Motl-Neit:2003}, we obtain the 
monodromy about the specified contour around the singularity to be
\be
\frac{A_+e^{5i\alpha_{+}}+A_-e^{5i\alpha_{-}}}
{A_+e^{i\alpha_{+}}+A_-e^{i\alpha_{-}}} \;.
\label{eq:monod1}
\ee
Eliminating the constants $A_{\pm}$ using the constraint (\ref{eq:constApm}), 
we get
\beq
-~\frac{e^{6i\alpha_{+}} - e^{6i\alpha_{-}}}
{e^{2i\alpha_{+}} - e^{2i\alpha_{-}}} = 
\frac{\sin(3\pi\nu)}{\sin (\pi\nu)}=  
1 + 2 \cos\l(\frac{\pi}{2} [D q - 2]\r) \, .
\eeq
We can evaluate the monodromy of $R(z)$ by choosing a contour which
passes through the horizon. The mode function near the horizon (as
$z\rightarrow -\infty$) is given by
\beq
R(z) \sim e^{+i\omega z} \sim \exp\left[i \, c_0 \, \omega\ln(x-x_{h})\right]
\;.
\label{eq:solr=rh}
\eeq
Thus, the monodromy by choosing the contour near the horizon is
$\exp(4 \pi \omega c_0)$. Equating the two monodromies, we obtain
\beq
\!\!\! \omega_{_{QNM}}  =  \frac{i}{2 c_0}\l[n + \frac{1}{2}\r] \pm 
\frac{1}{4 \pi c_0}\log\l[1 + 2 \cos\l(\frac{\pi}{2} [D q - 2]\r)\r] \, ,
\la{mainresult}
\eeq
where $n$ is an integer. This is the main result of our paper,
regarding which we would like to stress a few points. First, the above
result is valid for a general spherically symmetric space-times which
is asymptotically flat and has a single horizon at $r = r_h$. Second,
it is clear, from the above expression, that the imaginary part of the
high frequency QNM are discrete and are equally spaced. Third, the
real part of the high frequency QNM has a logarithmic dependence and
has a prefactor which depends on $r_h$ (See Table I). Lastly and more
importantly, it is clear from the above expression that the argument
of the logarithm is not always an integer. It is a non-negative
integer {\it only} if
\br
\label{eq:case-a}
&a) & \l(\frac{Dq - 2}{2}\r) \pi = 2 \, j \pi\quad {\rm where}\quad j \in
\mathcal{Z} \\ 
&b) & \l(\frac{Dq - 2}{2}\r) \pi = \frac{m \, \pi}{3}
\quad {\rm where}\quad m \in \mathcal{Z} \, .
\er
For the case (a), the real part of high frequency QNM is proportional
to $\log(3)$ for all $j$.  For the case (b), the real part of high
frequency QNM is (i) proportional to $\log(2)$ for $m = \pm 1, \pm 5,
\pm 7, \cdots$, and (ii) blows up for $m = \pm 2,
\pm 4, \pm 8, \cdots$. It is also interesting to note that the real
part of the QN frequencies {\it vanish} for all half odd integers in
Eq. (\ref{eq:case-a}).

\section{Application to specific black-holes}
\label{sec:specificBH}

In the previous section, we have obtained the high frequency QNM for a
spherically symmetric black hole with a generic singularity.  As we
have shown, the real part of the high frequency QNM is not necessarily
proportional to $\ln(3)$ as in the case of $(D + 2)$-dimensional
Schwarzschild. In order to illustrate this fact, we take specific
examples and obtain their QNM.

\subsection{$(D+2)-$dimensional Schwarzschild}

As noted earlier, for these black-holes
\be
4\p c_0 = \f{1}{T_H} ~~\mbox{and}~~
q = \f{2}{D}~.
\ee
Thus from (\ref{mainresult}), 
\beq
\omega_{_{QNM}}  =  2\p i T_H \l[n + \frac{1}{2}\r] \pm 
T_H \log 3~,
\label{eq:omeSch}
\eeq
as found by previous authors \cite{Motl-Neit:2003}. 

\subsection{$(D+2)-$dimensional extremal Reissner-Nordstr\"om}

{}From Table I, we see:
\be
c_0 = \f{D r_h}{(D-1)^2}~~\text{and}~~q=\f{2}{2D-1}~.
\ee
Thus 
\ba
\omega_{_{QNM}}  &=&  \f{i(D-1)^2}{2 D r_h} \l[n + \frac{1}{2}\r]  \nn \\
& \pm &\f{(D-1)^2}{4\p D r_h} 
 \log \le[ 1 + 2\cos\le( \f{\p(D-1)}{2D-1} \ri) \ri]~.
\ea
Thus, for example in four dimensions ($D=2$), we get:
\be
\omega_{_{QNM}}  =  \f{i}{4 r_h} \l[n + \frac{1}{2}\r] 
 \pm \f{1}{8\p r_h}  \log 2~~.
\label{eq:omeRN}
\ee
Note that the logarithmic nature of the real part of the QN frequency
persists although its coefficient is no longer the Hawking
temperature. The real part is in agreement with the argument of Motl
and Neitzke \cite{Motl-Neit:2003}, although their analysis is for
non-extremal Reissner-N\"ordstrom. Note however, that the above result 
appears to disagree with that stated in Ref. \cite{Cardoso-Nata:2004} and 
\cite{Natario-Schia:2004}.

\subsection{$2$-dimensional stringy black-holes}

Let us consider the generic $2$-dimensional black-hole solution in
string theory, which encompasses the solutions found in
Refs. \cite{Witten:1991,Mandal-Seng:1991,Callan-Gidd:1992}, with or
without matter fields
\footnote{The transformations 
$\eta = - \tanh^{-1} 
\le[ 1 - \f{M}{\lambda} e^{2\lambda r}\ri]^{-1}/\lambda$ 
and 
$x_\pm = \exp(-\lambda (r' \pm t))~({\rm where}~\exp(2\lambda r') 
= (-\lambda^2 \exp(2\lambda r))/(1 - \m \exp(2\lambda r)/\lambda)$
convert the metric (\ref{2dstring}) into the forms assumed in 
\cite{Witten:1991} and \cite{Callan-Gidd:1992} respectively.
}
:
\be 
ds^2 = - \le( 1 - \f{M}{\lambda}~e^{2\lambda r} \ri) dt^2 
+ \le( 1 - \f{M}{\lambda}~e^{2\lambda r} \ri)^{-1} dr^2 ~,
\la{2dstring}
\ee
where $\lambda$ is a constant and $M$ can be interpreted as the mass
of the black-hole. 
It has a horizon at:
\be
r = \f{1}{2 \lambda}\ln \le( \f{\lambda}{M} \ri) ~,
\ee
whose Hawking temperature is 
%
\be
T_H = \f{\lambda}{2\pi}~.
\ee
(Note that it is independent of $M$.) $D=0$ for these black-holes
renders $q$ irrelevant for QNM. QN frequencies can thus be read-off
from (\ref{mainresult}):
\beq
\o_{_{QNM}} = 2\p i T_H (n + m +1 ) 
~\mbox{or}~ 2\p i T_H (n - m ) ,~m \in {\mathcal Z}  \, .
\eeq
It is clear from the above expression that the real part of the high
QN frequencies for generic $2$-dimensional (stringy) black-holes
vanish.  In Ref. \cite{Kettner-Kuns:2004}, the authors have obtained
high-frequency QNM for a generic $2$-dimensional dilaton gravity. Our
result is in agreement with the results their case for $h(\phi) = 1$
which corresponds to pure $2$-dimensional gravity. 

\subsection{$4$-dimensional Stringy Black-Holes}

The line-element of the $4$-dimensional generalization of a charged
black-hole solution \cite{Garfinkle-Horo:1991,Gibbons-Maed:1987} is
given by (\ref{eq:spher-tr}) where
\ba
f(r)~&=&~g(r) = \l(1 - \frac{r_h}{r} \r)
\l(1 - \frac{r_0}{r}\r)^{(1-\a^2)/(1+\a^2)}; \nn \\
\rho(r) &=& r \l(1 - \frac{r_0}{r}\r)^{\a^2/(1+\a^2)} \, .
\ea
$r_h, r_0$ are related to the physical mass and charge by the 
relation
\beq
M = \frac{r_h}{2} + \l[\frac{1-\a^2}{1+\a^2}\r] \frac{r_0}{2};~~~~
Q = \l[\frac{r_h \, r_0}{1+\a^2}\r]^{1/2} \, .
\eeq
Note that $\a=0$ reduces it to the familiar $4$-dimensional
Reissner-Nordstr\"om solution, whereas for $\a=1$ the line-element
takes the form of the charged stringy black-hole solution of
\cite{Garfinkle-Horo:1991}.  The above solution has a regular
horizon at $r_h$. For any non-zero value of $\alpha$, the
inner-horizon $(r_0)$ is a space-like singularity. Thus, we will focus
on the situation where $\a \neq 0$ which corresponds to a black-hole
with a singular horizon.

It can be easily shown that the above black-hole has a non-zero
Hawking temperature given by:
\be
T_H = \f{1}{4\p r_h} \le(1 - \f{r_0}{r_h} \ri)^{(1-\a^2)/(1+\a^2)}~. 
\ee
Near the singularity $(r \to r_0)$, we have [$\e \equiv r - r_0$]:
\br
f(r) &=& \le( 1 - \f{r_h}{r_0} \ri)~
\le( \f{\e}{r_0} \ri)^{(1-\a^2)/( 1 + \a^2)} \, , \nn \\
\rho(r) &=& \le( r_0~\e^{\a^2} \ri)^{1/(1+\a^2)} \, ,\nn \\
x &=& \le( r_0~\e\ri)^{2\a^2/(1+\a^2)} \, .
\er
Thus, the $4$-dimensional stringy black-hole line-element near the 
singularity becomes
\br
\!\!\!\!\!\!\! 
ds^2 &\stackrel{r \to r_0}{\simeq}& h \, x^{(1 - \a^2)/(2 \a^2)} \!
\l[dt^2 - dx^2\r] + \, x \, d\Omega^2 \, ,
\er
where $h=h(r_0,r_h,\a)={\cal O}(1)$.
Comparing the above line-element with
(\ref{eq:singl-line2}), it follows that:
\beq
p = \frac{1 - \a^2}{2 \a^2};\quad q = 1 \, .
\eeq
Thus, from (\ref{mainresult}), we get:
\beq
\omega_{_{QNM}}  =  2\p i T_H \l[n + \frac{1}{2}\r] \pm 
T_H \log(3) ~.
\la{eq:omeGHS}
\ee
The above result is in agreement with Ref. \cite{Tamaki-Nomu:2004}.

\subsection{$5$-dimensional Stringy black-holes}

The line-element for RR charged $5$-dimensional stringy black-hole 
\cite{Horowitz-Mald:1996} 
formed by wrapping $D1$ and $D5$ branes on $T^4 \times S^1$ is given
by (\ref{eq:spher-tr}) where
\br
f(r) = F^{-2/3} \le( 1 - \f{r_h^2}{r^2} \ri);& & \rho(r) = F^{1/6} \,
r; \nn\\ g(r) = F^{-1/3} \le( 1 - \f{r_h^2}{r^2} \ri),& &
\er
and
\beq
F = \le[1 + \f{r_h^2 \sinh^2\a}{r^2} \ri] 
\le[ 1 + \f{r_h^2 \sinh^2\c}{r^2} \ri] 
\le[ 1 + \f{r_h^2 \sinh^2\s}{r^2} \ri] \nn  \, .
\eeq
The black-hole carries three $U(1)$ charges, which are proportional 
to the number of $D1-$branes, $D5-$branes and open string momentum 
along the compact dimension common between these branes, are related
to the black-hole parameters as:
\ba
Q_1 &=& \f{V r_h^2}{2g} \sinh 2\a ~;~ Q_5 = \f{ r_h^2}{2g} \sinh 2\c ; \nn \\ 
n &=& \f{ R^2 V r_h^2}{2g^2} \sinh 2\s~,
\ea
where $(2\p)^4 V$ and $2\p R$ are the volume and radius of the $T^4$
and $S^1$ respectively and $g$ is the string coupling. The above
solution has a regular event horizon at $r_h$. When all the three
charges are nonzero, the surface $r = 0$ is a smooth inner
horizon. When at least one of the charges is zero (say $\sigma = 0$),
the event horizon remains, however the surface $r = 0$ becomes
singular. Thus, we will focus on the situation where $\sigma = 0$
which corresponds to a black-hole with a singularity at $r = 0$.
The surface gravity for this resultant black-hole is
\beq
\kappa = \l(r_h \, \cosh\a \, \cosh\c \r)^{-1} \, .
\eeq
Near the singularity, we have
\br
f(r)= -k~r^{2/3}\, ;& &
g(r) = -l~r^{2/3}\, ; \nn \\
\rho(r) = m~r^{1/3}\, ; & & 
x = \le( m \sqrt{kl} \ri)~r \, .
\er
Thus, the $5$-dimensional stringy black-hole line-element near the 
singularity becomes
\br
\!\!\!\!\!\!\! 
ds^2 &\stackrel{r \to 0}{\simeq}& s \, x^{2/3} \!
\l[dt^2 - dx^2\r] +  x^{2/3} \, d\Omega^2 
\er
where $k,l,m,s = k,l,m,s~(r_h,\a,\gamma) = {\cal O}(1)$. 
Comparing with (\ref{eq:singl-line2}), it follows that:
\beq
p = \frac{2}{3};\quad q = \frac{2}{3} \, .
\eeq
Thus, from (\ref{mainresult}), we get:
\beq
\omega_{_{QNM}}  =  2\p i T_H \l[n + \frac{1}{2}\r] \pm 
T_H \log(3) ~.
\la{eq:omeHMS}
\ee
We would like to clarify the following point regarding our result: As
we had mentioned earlier, the analysis is {\it strictly} valid for
single-horizon black-hole space-times. In obtaining the high-frequency
QNMs for the 5-dimensional stringy black-holes, we have assumed that
one of the charges to the zero ($\sigma = 0$). Note, however, for
these stringy black-holes with a regular inner horizon, it has been
shown in Ref. \cite{Birmingham-Carl:2003} that the frequencies
determined by the monodromies (of the two horizons) coincide with the
QNM of the near horizon BTZ metric and that these are the ones that
are relevant for quantum gravity (the so called `non-quasinormal
modes'). It will be interesting to closely examine the relation
between the two results.

%
%
\begin{table*}[!hbt]
\caption{The table gives the list of physical quantities for different 
black-holes. $k$ is the surface gravity of the black-hole; $c_0$ is
the constant which appears in the near-horizon relation between $x$
and $r$; $p$ and $q$ are the power-law index for ${\cal M}^2$ and $S^D$
respectively, and $(D \,q - 2)/2$ is the quantity which determines the
real part of QN frequency $(\Re[\omega_{_{QNM}}])$. The properties of the 
various black-holes (BH) are discussed in Secs. (\ref{sec:SphSy}, 
\ref{sec:specificBH}).}
\label{ta:1}
\begin{tabular}{|c||c|c||c|c|c||c|}
\hline
Space-Time&\multicolumn{2}{c|}{Near horizon properties}&\multicolumn{3}{c|}{Near singularity properties} & $\Re[\omega_{_{\rm QNM}}]$\\ 
\cline{2-6}
 &$\kappa$&$c_0$&$p$&$q$&$ (D q - 2)/2$ & \\ 
\hline\hline 
Non-degenerate&$\frac{\sqrt{f'(r_h) g'(r_h)}}{2}$& 
$\frac{1}{2 \kappa} = \frac{1}{4 \pi T_H}$&$p > q - 2$&$q > 0$ & 
Any real value & 
$\frac{1}{4 \pi c_0}\log\l[1 + 2 \cos\l(\frac{\pi (D q - 2)}{2}\r)\r]$ \\
Horizons& & & & & &  \\
& & & & & & \\
(D + 2)-dimens. &$\frac{(D - 1)}{2 r_h}$ & 
$\frac{r_h}{D - 1}$ & $\frac{1 - D}{D}$ &  $\frac{2}{D}$ & 0 & 
$T_H \log 3$ \\
Schwarzchild & & & & & & \\
& & & & & & \\
4-dimensional &$\f{1}{2 r_h}\le[1 - \f{r_0}{r_h}
\ri]^{\frac{1-\a^2}{1+\a^2}}$ & $ \frac{1}{2 \kappa}$ & $\frac{1 -
\a^2}{2 \a^2}$ & $1$ & $0$ & $T_H \log 3$ \\
stringy BH & & & & & & \\
& & & & & & \\
5-dimensional &$\f{r_h^{-1}}{\cosh\a \cosh\c}$ & 
$ \frac{1}{2 \kappa}$ & $-\f{2}{3}$ & $\f{2}{3}$ & $0$ & $T_H \log 3$ \\
stringy BH & & & & & & \\
& & & & & & \\
2-dimensional &$1$ & 
$ \frac{1}{2}$ & $1$ & $0$ & $-1$ & $0$ \\
stringy BH & & & & & & \\
& & & & & & \\
(D + 2)-dimens. & $0$ & 
$\frac{D}{(D - 1)^2} r_h$ & 
$- 2 \le( \f{D-1}{2D-1} \ri)$ & $\f{2}{2D-1}$ & 
$\l(\frac{1 - D}{2 D - 1}\r)$ & $\f{1}{8\p r_h}  \log 2$ \\
Degenerate RN&  & & & & & \\
\hline
\end{tabular}
\end{table*}
	
\section{Discussion}
\label{sec:conc}

In this paper, we have computed the high frequency QNM for scalar
perturbations of spherically symmetric single horizon and
asymptotically flat black-holes in $(D+2)-$dimensional space-times. We
have computed these modes using the monodromy approach
\cite{Motl-Neit:2003}. We have shown that the real part of these modes 
depends on the horizon radius ($r_h$), dimension ($D$) and the
power-law index ($q$) of $S^D$ near the singularity.

We have also shown that the real part of the high frequency modes has
a logarithmic dependence, although the argument of the logarithm is
not necessarily an integer. However, when we applied our result to
specific examples and obtained their QNM, the argument turned out to
be an integer. In particular, for $(D + 2)$-dimensional Schwarzschild
and $4,5$-dimensional stringy black-holes, we found that the real part
of the QN frequencies is proportional to $\ln(3)$. However for
$4$-dimensional Reissner-Nordstr\"om it is proportional to $\ln(2)$
and vanishes in the case of $2$-dimensional stringy black-holes.  It
would be interesting to compute the QNM for black holes with multiple
horizons as well as those which are not spherically symmetric or
asymptotically flat and also to compare our results with others that
are related to a generic class of black holes
\cite{Kettner-Kuns:2004}.

In the light of the above results, let us re-examine the argument
presented in \cite{Dreyer:2002}. For black-holes with non-zero
Hawking temperature, it follows from (\ref{mainresult}) that the real
part of QN frequencies is of the form:
\be
\Re(\o_{QNM}) = T_H \log \le[ 1 + 2 \cos \le( \f{\p(Dq-2)}{2} \ri) \ri]~.
\label{eq:ReOm}
\ee 
Identifying the above with the minimum change of black-hole mass
(due to Hawking radiation or accretion), we get, using the first law
of black-hole thermodynamics and the area law for entropy:
\be
\Delta A = 4 \lPl^2 
\log \le[ 1 + 2 \cos \le( \f{\p(Dq-2)}{2} \ri) \ri]~,
\label{eq:DArea}
\ee
where $A=\O_D r_h^D$ is the horizon area of the black-hole and 
$\lPl$ is the $(D+2)-$dimensional Planck length in this case. 
Equating with the minimum area quantum in loop quantum gravity, namely
$\Delta A = 8\p \lPl^2 \gamma \sqrt{j_m(j_m+1)}$ and subsequently
setting $D=2$, we get the following prediction for the Immirzi parameter:
\be
\gamma = \f{
\log \le[ 1 + 2 \cos \le( \p(q-1) \ri) \ri]}{2\p \sqrt{j_m(j_m+1)}}~,
\ee
where $j_m$ denotes the representation of $SU(2)$ for the spin-network
states in loop quantum gravity (for super-symmetric spin networks the
area spectrum is derived in Ref. \cite{Ling-Zhan:2003}). Assuming that
the number of points where the spin-network states puncture the
horizon is given by:
\be
N = A/\Delta A 
\ee
one gets the microscopic entropy of the black-hole, as the logarithm of
the dimension of the boundary Hilbert space, to be
\ba
S &=& N \log (1+ 2j_m)  \nn \\ 
&=& \f{A}{4\lPl^2} \f{\ln \le(1 + 2 j_m \ri) }
{\log \le[ 1 + 2 \cos \le( \p(q-1) \ri) \ri]}
\ea
Thus, for the above relation to agree with the Bekenstein-Hawking entropy,
one must have:
\be
j_m = \cos\le( \p(q-1) \ri)~.
\label{eq:jm}
\ee
Even though, for $q = 1$, (\ref{eq:jm}) agrees with the prediction
of Ref. \cite{Dreyer:2002}, one would like to have a better
understanding of the above result. More importantly, the above result
is valid {\it only} for non-extremal black-holes. Since $\kappa = 0$
for the extremal black-holes, the real part of the QN frequencies
(\ref{eq:ReOm}) cannot be related to the temperature implying that the
relation for the change in the horizon area (\ref{eq:DArea}) is no
more valid. We hope to address this issue elsewhere.

Finally, following \cite{Kunstatter:2002} we find that the adiabatic
invariant in our case is:
\ba
I &=& \int \f{dE}{\o_{QNM}} \nn \\
&=& \l(\log \le[ 1 + 2 \cos \le( \f{\p(Dq-2)}{2} \ri) \ri]\r)^{-1}~ 
\int \f{dE}{T_H}  \nn \\
&=& \l(\log \le[ 1 + 2 \cos \le( \f{\p(Dq-2)}{2} \ri) \ri]\r)^{-1} S~,
\ea
which again confirms Bekenstein's conjecture that horizon area (and
hence black-hole entropy) is an adiabatic invariant.  The crucial
ingredient in the above is:
\be
\Re(\o_{QNM}) \propto T_H~.
\ee
This, along with the Bohr-Sommerfeld quantization rule $I = n$ 
implies the equispaced nature of the horizon spectrum:
\be
A \propto n~\lPl^D~,
\ee
with the following proportionality constant:
\be
4 \times \l(\log \le[ 1 + 2 \cos \le( \f{\p(Dq-2)}{2} \ri) \ri]\r)^{-1}
\ee
Such equispaced spectrum has been verified earlier in several other
approaches, albeit with different proportionality constants and often
with a `zero-point' or `ground state' energy, interpreted as a
Planck-sized remnant left over when the black-hole evaporates (see for
e.g. \cite{Barvinsky-Das:2002,Das-Rama:2002,Dasgupta:2003} and
references therein). It is also interesting to note from
(\ref{eq:omeRN}) and the relation $r_h \sim M^{1/(D-1)}$ that the
relation $I \propto A$ continues to hold for extremal
Reissner-N\"ordstrom.  

We end with a couple of caveats: (i) The area spectrum in Loop Quantum
Gravity is not equispaced in general
\cite{Rovelli-Smol:1994,Ashtekar-Lewa:1996}. However, it is equispaced
in the large area limit, as well as if one uses a different
representation
\cite{Alekseev-Poly:2000}.  (ii) Strictly speaking, the QNM are only
associated with transitions of short durations. Thus, other
transitions can modify the equispaced area spectrum obtained here.
We hope to further examine these and related issues elsewhere.

\section*{Acknowledgments}

This work was supported by the Natural Sciences and Engineering
Research Council of Canada. We would like to thank 
G. Kunstatter, A. J. M. Medved, 
J. M. Natario and R. Schiappa for useful comments.  
SS would like to thank the Department of
Physics, University of Lethbridge, Canada for hospitality where this
work was completed. SS would also like to thank Martin O'Loughlin and
F. Hussain for discussions.


\begin{thebibliography}{61}
\expandafter\ifx\csname natexlab\endcsname\relax\def\natexlab#1{#1}\fi
\expandafter\ifx\csname bibnamefont\endcsname\relax
  \def\bibnamefont#1{#1}\fi
\expandafter\ifx\csname bibfnamefont\endcsname\relax
  \def\bibfnamefont#1{#1}\fi
\expandafter\ifx\csname citenamefont\endcsname\relax
  \def\citenamefont#1{#1}\fi
\expandafter\ifx\csname url\endcsname\relax
  \def\url#1{\texttt{#1}}\fi
\expandafter\ifx\csname urlprefix\endcsname\relax\def\urlprefix{URL }\fi
\providecommand{\bibinfo}[2]{#2}
\providecommand{\eprint}[2][]{\url{#2}}

\bibitem[{\citenamefont{Kokkotas and Schmidt}(1999)}]{Kokkotas-Schm:1999}
\bibinfo{author}{\bibfnamefont{K.~D.} \bibnamefont{Kokkotas}} \bibnamefont{and}
  \bibinfo{author}{\bibfnamefont{B.~G.} \bibnamefont{Schmidt}},
  \bibinfo{journal}{Living Rev. Rel.} \textbf{\bibinfo{volume}{2}},
  \bibinfo{pages}{2} (\bibinfo{year}{1999}), \eprint{gr-qc/9909058}.

\bibitem[{\citenamefont{Horowitz and Hubeny}(2000)}]{Horowitz-Hube:1999}
\bibinfo{author}{\bibfnamefont{G.~T.} \bibnamefont{Horowitz}} \bibnamefont{and}
  \bibinfo{author}{\bibfnamefont{V.~E.} \bibnamefont{Hubeny}},
  \bibinfo{journal}{Phys. Rev.} \textbf{\bibinfo{volume}{D62}},
  \bibinfo{pages}{024027} (\bibinfo{year}{2000}), \eprint{hep-th/9909056}.

\bibitem[{\citenamefont{Hod}(1998)}]{Hod:1998}
\bibinfo{author}{\bibfnamefont{S.}~\bibnamefont{Hod}}, \bibinfo{journal}{Phys.
  Rev. Lett.} \textbf{\bibinfo{volume}{81}}, \bibinfo{pages}{4293}
  (\bibinfo{year}{1998}), \eprint{gr-qc/9812002}.

\bibitem[{\citenamefont{Dreyer}(2003)}]{Dreyer:2002}
\bibinfo{author}{\bibfnamefont{O.}~\bibnamefont{Dreyer}},
  \bibinfo{journal}{Phys. Rev. Lett.} \textbf{\bibinfo{volume}{90}},
  \bibinfo{pages}{081301} (\bibinfo{year}{2003}), \eprint{gr-qc/0211076}.

\bibitem[{\citenamefont{Bekenstein}(1974)}]{Bekenstein:1974}
\bibinfo{author}{\bibfnamefont{J.~D.} \bibnamefont{Bekenstein}},
  \bibinfo{journal}{Lett. Nuovo Cim.} \textbf{\bibinfo{volume}{11}},
  \bibinfo{pages}{467} (\bibinfo{year}{1974}).

\bibitem[{\citenamefont{Bekenstein and Mukhanov}(1995)}]{Bekenstein-Mukh:1995}
\bibinfo{author}{\bibfnamefont{J.~D.} \bibnamefont{Bekenstein}}
  \bibnamefont{and} \bibinfo{author}{\bibfnamefont{V.~F.}
  \bibnamefont{Mukhanov}}, \bibinfo{journal}{Phys. Lett.}
  \textbf{\bibinfo{volume}{B360}}, \bibinfo{pages}{7} (\bibinfo{year}{1995}),
  \eprint{gr-qc/9505012}.

\bibitem[{\citenamefont{Nollert}(1993)}]{Nollert:1993}
\bibinfo{author}{\bibfnamefont{H.-P.} \bibnamefont{Nollert}},
  \bibinfo{journal}{Phys. Rev.} \textbf{\bibinfo{volume}{D47}},
  \bibinfo{pages}{5253} (\bibinfo{year}{1993}).

\bibitem[{\citenamefont{Motl}(2003)}]{Motl:2002}
\bibinfo{author}{\bibfnamefont{L.}~\bibnamefont{Motl}}, \bibinfo{journal}{Adv.
  Theor. Math. Phys.} \textbf{\bibinfo{volume}{6}}, \bibinfo{pages}{1135}
  (\bibinfo{year}{2003}), \eprint{gr-qc/0212096}.

\bibitem[{\citenamefont{Domagala and Lewandowski}(2004)}]{Domagala-Lewa:2004}
\bibinfo{author}{\bibfnamefont{M.}~\bibnamefont{Domagala}} \bibnamefont{and}
  \bibinfo{author}{\bibfnamefont{J.}~\bibnamefont{Lewandowski}},
  \bibinfo{journal}{Class. Quant. Grav.} \textbf{\bibinfo{volume}{21}},
  \bibinfo{pages}{5233} (\bibinfo{year}{2004}), \eprint{gr-qc/0407051}.

\bibitem[{\citenamefont{Meissner}(2004)}]{Meissner:2004}
\bibinfo{author}{\bibfnamefont{K.~A.} \bibnamefont{Meissner}},
  \bibinfo{journal}{Class. Quant. Grav.} \textbf{\bibinfo{volume}{21}},
  \bibinfo{pages}{5245} (\bibinfo{year}{2004}), \eprint{gr-qc/0407052}.

\bibitem[{\citenamefont{Dreyer et~al.}(2004)\citenamefont{Dreyer, Markopoulou,
  and Smolin}}]{Dreyer-Mark:2004}
\bibinfo{author}{\bibfnamefont{O.}~\bibnamefont{Dreyer}},
  \bibinfo{author}{\bibfnamefont{F.}~\bibnamefont{Markopoulou}},
  \bibnamefont{and} \bibinfo{author}{\bibfnamefont{L.}~\bibnamefont{Smolin}}
  (\bibinfo{year}{2004}), \eprint{hep-th/0409056}.

\bibitem[{\citenamefont{Alexandrov}(2004)}]{Alexandrov:2004}
\bibinfo{author}{\bibfnamefont{S.}~\bibnamefont{Alexandrov}}
  (\bibinfo{year}{2004}), \eprint{gr-qc/0408033}.

\bibitem[{\citenamefont{Kunstatter}(2003)}]{Kunstatter:2002}
\bibinfo{author}{\bibfnamefont{G.}~\bibnamefont{Kunstatter}},
  \bibinfo{journal}{Phys. Rev. Lett.} \textbf{\bibinfo{volume}{90}},
  \bibinfo{pages}{161301} (\bibinfo{year}{2003}), \eprint{gr-qc/0212014}.

\bibitem[{\citenamefont{Motl and Neitzke}(2003)}]{Motl-Neit:2003}
\bibinfo{author}{\bibfnamefont{L.}~\bibnamefont{Motl}} \bibnamefont{and}
  \bibinfo{author}{\bibfnamefont{A.}~\bibnamefont{Neitzke}},
  \bibinfo{journal}{Adv. Theor. Math. Phys.} \textbf{\bibinfo{volume}{7}},
  \bibinfo{pages}{307} (\bibinfo{year}{2003}), \eprint{hep-th/0301173}.

\bibitem[{\citenamefont{Andersson and Howls}(2004)}]{Andersson-Howl:2003}
\bibinfo{author}{\bibfnamefont{N.}~\bibnamefont{Andersson}} \bibnamefont{and}
  \bibinfo{author}{\bibfnamefont{C.~J.} \bibnamefont{Howls}},
  \bibinfo{journal}{Class. Quant. Grav.} \textbf{\bibinfo{volume}{21}},
  \bibinfo{pages}{1623} (\bibinfo{year}{2004}), \eprint{gr-qc/0307020}.

\bibitem[{\citenamefont{Suneeta}(2003)}]{Suneeta:2003}
\bibinfo{author}{\bibfnamefont{V.}~\bibnamefont{Suneeta}},
  \bibinfo{journal}{Phys. Rev.} \textbf{\bibinfo{volume}{D68}},
  \bibinfo{pages}{024020} (\bibinfo{year}{2003}), \eprint{gr-qc/0303114}.

\bibitem[{\citenamefont{Medved et~al.}(2004)\citenamefont{Medved, Martin, and
  Visser}}]{Medved-Martin:2003}
\bibinfo{author}{\bibfnamefont{A.~J.~M.} \bibnamefont{Medved}},
  \bibinfo{author}{\bibfnamefont{D.}~\bibnamefont{Martin}}, \bibnamefont{and}
  \bibinfo{author}{\bibfnamefont{M.}~\bibnamefont{Visser}},
  \bibinfo{journal}{Class. Quant. Grav.} \textbf{\bibinfo{volume}{21}},
  \bibinfo{pages}{1393} (\bibinfo{year}{2004}), \eprint{gr-qc/0310009}.

\bibitem[{\citenamefont{Padmanabhan}(2004)}]{Paddy:2003}
\bibinfo{author}{\bibfnamefont{T.}~\bibnamefont{Padmanabhan}},
  \bibinfo{journal}{Class. Quant. Grav.} \textbf{\bibinfo{volume}{21}},
  \bibinfo{pages}{L1} (\bibinfo{year}{2004}), \eprint{gr-qc/0310027}.

\bibitem[{\citenamefont{Choudhury and Padmanabhan}(2004)}]{Tirth-Paddy:2003}
\bibinfo{author}{\bibfnamefont{T.~R.} \bibnamefont{Choudhury}}
  \bibnamefont{and}
  \bibinfo{author}{\bibfnamefont{T.}~\bibnamefont{Padmanabhan}},
  \bibinfo{journal}{Phys. Rev.} \textbf{\bibinfo{volume}{D69}},
  \bibinfo{pages}{064033} (\bibinfo{year}{2004}), \eprint{gr-qc/0311064}.

\bibitem[{\citenamefont{Cardoso}(2003)}]{Cardoso-Phd:2003}
\bibinfo{author}{\bibfnamefont{V.}~\bibnamefont{Cardoso}}
  (\bibinfo{year}{2003}), \eprint{gr-qc/0404093}.

\bibitem[{\citenamefont{Kettner et~al.}(2004)\citenamefont{Kettner, Kunstatter,
  and Medved}}]{Kettner-Kuns:2004}
\bibinfo{author}{\bibfnamefont{J.}~\bibnamefont{Kettner}},
  \bibinfo{author}{\bibfnamefont{G.}~\bibnamefont{Kunstatter}},
  \bibnamefont{and} \bibinfo{author}{\bibfnamefont{A.~J.~M.}
  \bibnamefont{Medved}} (\bibinfo{year}{2004}), \eprint{gr-qc/0408042}.

\bibitem[{\citenamefont{Govindarajan and
  Suneeta}(2001)}]{Govindarajan-Sune:2000}
\bibinfo{author}{\bibfnamefont{T.~R.} \bibnamefont{Govindarajan}}
  \bibnamefont{and} \bibinfo{author}{\bibfnamefont{V.}~\bibnamefont{Suneeta}},
  \bibinfo{journal}{Class. Quant. Grav.} \textbf{\bibinfo{volume}{18}},
  \bibinfo{pages}{265} (\bibinfo{year}{2001}), \eprint{gr-qc/0007084}.

\bibitem[{\citenamefont{Birmingham}(2003)}]{Birmingham:2003}
\bibinfo{author}{\bibfnamefont{D.}~\bibnamefont{Birmingham}},
  \bibinfo{journal}{Phys. Lett.} \textbf{\bibinfo{volume}{B569}},
  \bibinfo{pages}{199} (\bibinfo{year}{2003}), \eprint{hep-th/0306004}.

\bibitem[{\citenamefont{Birmingham et~al.}(2003)\citenamefont{Birmingham,
  Carlip, and Chen}}]{Birmingham-Carp:2003}
\bibinfo{author}{\bibfnamefont{D.}~\bibnamefont{Birmingham}},
  \bibinfo{author}{\bibfnamefont{S.}~\bibnamefont{Carlip}}, \bibnamefont{and}
  \bibinfo{author}{\bibfnamefont{Y.-j.} \bibnamefont{Chen}},
  \bibinfo{journal}{Class. Quant. Grav.} \textbf{\bibinfo{volume}{20}},
  \bibinfo{pages}{L239} (\bibinfo{year}{2003}), \eprint{hep-th/0305113}.

\bibitem[{\citenamefont{Birmingham et~al.}(2002)\citenamefont{Birmingham,
  Sachs, and Solodukhin}}]{Birmingham-Sach:2001}
\bibinfo{author}{\bibfnamefont{D.}~\bibnamefont{Birmingham}},
  \bibinfo{author}{\bibfnamefont{I.}~\bibnamefont{Sachs}}, \bibnamefont{and}
  \bibinfo{author}{\bibfnamefont{S.~N.} \bibnamefont{Solodukhin}},
  \bibinfo{journal}{Phys. Rev. Lett.} \textbf{\bibinfo{volume}{88}},
  \bibinfo{pages}{151301} (\bibinfo{year}{2002}), \eprint{hep-th/0112055}.

\bibitem[{\citenamefont{Berti et~al.}(2004)\citenamefont{Berti, Cardoso, and
  Yoshida}}]{Berti-Cardoso:2004}
\bibinfo{author}{\bibfnamefont{E.}~\bibnamefont{Berti}},
  \bibinfo{author}{\bibfnamefont{V.}~\bibnamefont{Cardoso}}, \bibnamefont{and}
  \bibinfo{author}{\bibfnamefont{S.}~\bibnamefont{Yoshida}},
  \bibinfo{journal}{Phys. Rev.} \textbf{\bibinfo{volume}{D69}},
  \bibinfo{pages}{124018} (\bibinfo{year}{2004}), \eprint{gr-qc/0401052}.

\bibitem[{\citenamefont{Maassen van~den Brink}(2004)}]{MaassenvandenBrink:2003}
\bibinfo{author}{\bibfnamefont{A.}~\bibnamefont{Maassen van~den Brink}},
  \bibinfo{journal}{J. Math. Phys.} \textbf{\bibinfo{volume}{45}},
  \bibinfo{pages}{327} (\bibinfo{year}{2004}), \eprint{gr-qc/0303095}.

\bibitem[{\citenamefont{Musiri and Siopsis}(2003)}]{Musiri-Siopsis:2003c}
\bibinfo{author}{\bibfnamefont{S.}~\bibnamefont{Musiri}} \bibnamefont{and}
  \bibinfo{author}{\bibfnamefont{G.}~\bibnamefont{Siopsis}},
  \bibinfo{journal}{Phys. Lett.} \textbf{\bibinfo{volume}{B563}},
  \bibinfo{pages}{102} (\bibinfo{year}{2003}), \eprint{hep-th/0301081}.

\bibitem[{\citenamefont{Oppenheim}(2004)}]{Oppenheim:2003}
\bibinfo{author}{\bibfnamefont{J.}~\bibnamefont{Oppenheim}},
  \bibinfo{journal}{Phys. Rev.} \textbf{\bibinfo{volume}{D69}},
  \bibinfo{pages}{044012} (\bibinfo{year}{2004}), \eprint{gr-qc/0307089}.

\bibitem[{\citenamefont{Cardoso et~al.}(2003)\citenamefont{Cardoso, Konoplya,
  and Lemos}}]{Cardoso-Kono:2003}
\bibinfo{author}{\bibfnamefont{V.}~\bibnamefont{Cardoso}},
  \bibinfo{author}{\bibfnamefont{R.}~\bibnamefont{Konoplya}}, \bibnamefont{and}
  \bibinfo{author}{\bibfnamefont{J.~P.~S.} \bibnamefont{Lemos}},
  \bibinfo{journal}{Phys. Rev.} \textbf{\bibinfo{volume}{D68}},
  \bibinfo{pages}{044024} (\bibinfo{year}{2003}), \eprint{gr-qc/0305037}.

\bibitem[{\citenamefont{Cardoso
  et~al.}(2004{\natexlab{a}})\citenamefont{Cardoso, Lemos, and
  Yoshida}}]{Cardoso-Lemo:2003}
\bibinfo{author}{\bibfnamefont{V.}~\bibnamefont{Cardoso}},
  \bibinfo{author}{\bibfnamefont{J.~P.~S.} \bibnamefont{Lemos}},
  \bibnamefont{and} \bibinfo{author}{\bibfnamefont{S.}~\bibnamefont{Yoshida}},
  \bibinfo{journal}{Phys. Rev.} \textbf{\bibinfo{volume}{D69}},
  \bibinfo{pages}{044004} (\bibinfo{year}{2004}{\natexlab{a}}),
  \eprint{gr-qc/0309112}.

\bibitem[{\citenamefont{Konoplya}(2002)}]{Konoplya:2002}
\bibinfo{author}{\bibfnamefont{R.~A.} \bibnamefont{Konoplya}},
  \bibinfo{journal}{Phys. Rev.} \textbf{\bibinfo{volume}{D66}},
  \bibinfo{pages}{044009} (\bibinfo{year}{2002}), \eprint{hep-th/0205142}.

\bibitem[{\citenamefont{Konoplya}(2003)}]{Konoplya:2003}
\bibinfo{author}{\bibfnamefont{R.~A.} \bibnamefont{Konoplya}},
  \bibinfo{journal}{Phys. Rev.} \textbf{\bibinfo{volume}{D68}},
  \bibinfo{pages}{024018} (\bibinfo{year}{2003}), \eprint{gr-qc/0303052}.

\bibitem[{\citenamefont{Setare}(2004{\natexlab{a}})}]{Setare:2003}
\bibinfo{author}{\bibfnamefont{M.~R.} \bibnamefont{Setare}},
  \bibinfo{journal}{Class. Quant. Grav.} \textbf{\bibinfo{volume}{21}},
  \bibinfo{pages}{1453} (\bibinfo{year}{2004}{\natexlab{a}}),
  \eprint{hep-th/0311221}.

\bibitem[{\citenamefont{Konoplya}(2004)}]{Konoplya:2004}
\bibinfo{author}{\bibfnamefont{R.}~\bibnamefont{Konoplya}}
  (\bibinfo{year}{2004}), \eprint{hep-th/0410057}.

\bibitem[{\citenamefont{Setare}(2004{\natexlab{b}})}]{Setare:2004}
\bibinfo{author}{\bibfnamefont{M.~R.} \bibnamefont{Setare}},
  \bibinfo{journal}{Phys. Rev.} \textbf{\bibinfo{volume}{D69}},
  \bibinfo{pages}{044016} (\bibinfo{year}{2004}{\natexlab{b}}),
  \eprint{hep-th/0312061}.

\bibitem[{\citenamefont{Tamaki and Nomura}(2004)}]{Tamaki-Nomu:2004}
\bibinfo{author}{\bibfnamefont{T.}~\bibnamefont{Tamaki}} \bibnamefont{and}
  \bibinfo{author}{\bibfnamefont{H.}~\bibnamefont{Nomura}},
  \bibinfo{journal}{Phys. Rev.} \textbf{\bibinfo{volume}{D70}},
  \bibinfo{pages}{044041} (\bibinfo{year}{2004}), \eprint{hep-th/0405191}.

\bibitem[{\citenamefont{Szekeres and Iyer}(1993)}]{Szekeres-Iyer:1993}
\bibinfo{author}{\bibfnamefont{P.}~\bibnamefont{Szekeres}} \bibnamefont{and}
  \bibinfo{author}{\bibfnamefont{V.}~\bibnamefont{Iyer}},
  \bibinfo{journal}{Phys. Rev.} \textbf{\bibinfo{volume}{D47}},
  \bibinfo{pages}{4362} (\bibinfo{year}{1993}).

\bibitem[{\citenamefont{Celerier and Szekeres}(2002)}]{Celerier-Szek:2002}
\bibinfo{author}{\bibfnamefont{M.-N.} \bibnamefont{Celerier}} \bibnamefont{and}
  \bibinfo{author}{\bibfnamefont{P.}~\bibnamefont{Szekeres}},
  \bibinfo{journal}{Phys. Rev.} \textbf{\bibinfo{volume}{D65}},
  \bibinfo{pages}{123516} (\bibinfo{year}{2002}), \eprint{gr-qc/0203094}.

\bibitem[{\citenamefont{Blau et~al.}(2004{\natexlab{a}})\citenamefont{Blau,
  Borunda, O'Loughlin, and Papadopoulos}}]{Blau-Boru:2004}
\bibinfo{author}{\bibfnamefont{M.}~\bibnamefont{Blau}},
  \bibinfo{author}{\bibfnamefont{M.}~\bibnamefont{Borunda}},
  \bibinfo{author}{\bibfnamefont{M.}~\bibnamefont{O'Loughlin}},
  \bibnamefont{and}
  \bibinfo{author}{\bibfnamefont{G.}~\bibnamefont{Papadopoulos}},
  \bibinfo{journal}{JHEP} \textbf{\bibinfo{volume}{07}}, \bibinfo{pages}{068}
  (\bibinfo{year}{2004}{\natexlab{a}}), \eprint{hep-th/0403252}.

\bibitem[{\citenamefont{Wald}(1984)}]{Wald:1984-bk}
\bibinfo{author}{\bibfnamefont{R.~M.} \bibnamefont{Wald}},
  \emph{\bibinfo{title}{General Relativity}} (\bibinfo{publisher}{Chicago Univ.
  Pr., USA}, \bibinfo{year}{1984}).

\bibitem[{\citenamefont{Gibbons and Maeda}(1988)}]{Gibbons-Maed:1987}
\bibinfo{author}{\bibfnamefont{G.~W.} \bibnamefont{Gibbons}} \bibnamefont{and}
  \bibinfo{author}{\bibfnamefont{K.}~\bibnamefont{Maeda}},
  \bibinfo{journal}{Nucl. Phys.} \textbf{\bibinfo{volume}{B298}},
  \bibinfo{pages}{741} (\bibinfo{year}{1988}).

\bibitem[{\citenamefont{Garfinkle et~al.}(1991)\citenamefont{Garfinkle,
  Horowitz, and Strominger}}]{Garfinkle-Horo:1991}
\bibinfo{author}{\bibfnamefont{D.}~\bibnamefont{Garfinkle}},
  \bibinfo{author}{\bibfnamefont{G.~T.} \bibnamefont{Horowitz}},
  \bibnamefont{and}
  \bibinfo{author}{\bibfnamefont{A.}~\bibnamefont{Strominger}},
  \bibinfo{journal}{Phys. Rev.} \textbf{\bibinfo{volume}{D43}},
  \bibinfo{pages}{3140} (\bibinfo{year}{1991}).

\bibitem[{\citenamefont{Horowitz et~al.}(1996)\citenamefont{Horowitz,
  Maldacena, and Strominger}}]{Horowitz-Mald:1996}
\bibinfo{author}{\bibfnamefont{G.~T.} \bibnamefont{Horowitz}},
  \bibinfo{author}{\bibfnamefont{J.~M.} \bibnamefont{Maldacena}},
  \bibnamefont{and}
  \bibinfo{author}{\bibfnamefont{A.}~\bibnamefont{Strominger}},
  \bibinfo{journal}{Phys. Lett.} \textbf{\bibinfo{volume}{B383}},
  \bibinfo{pages}{151} (\bibinfo{year}{1996}), \eprint{hep-th/9603109}.

\bibitem[{\citenamefont{Kodama and Ishibashi}(2004)}]{Kodama-Ishi:2003}
\bibinfo{author}{\bibfnamefont{H.}~\bibnamefont{Kodama}} \bibnamefont{and}
  \bibinfo{author}{\bibfnamefont{A.}~\bibnamefont{Ishibashi}},
  \bibinfo{journal}{Prog. Theor. Phys.} \textbf{\bibinfo{volume}{111}},
  \bibinfo{pages}{29} (\bibinfo{year}{2004}), \eprint{hep-th/0308128}.

\bibitem[{\citenamefont{Morse and Feshbach}(1953)}]{Morse-Fesh:1953-bk1}
\bibinfo{author}{\bibfnamefont{P.~M.} \bibnamefont{Morse}} \bibnamefont{and}
  \bibinfo{author}{\bibfnamefont{H.}~\bibnamefont{Feshbach}},
  \emph{\bibinfo{title}{Methods of Theoretical Physics, Part I}}
  (\bibinfo{publisher}{McGraw-Hill, New York}, \bibinfo{year}{1953}).

\bibitem[{\citenamefont{Blau and O'Loughlin}(2004)}]{Blau-OLou:2004}
\bibinfo{author}{\bibfnamefont{M.}~\bibnamefont{Blau}} \bibnamefont{and}
  \bibinfo{author}{\bibfnamefont{M.}~\bibnamefont{O'Loughlin}},
  \bibinfo{journal}{private communication}  (\bibinfo{year}{2004}).

\bibitem[{\citenamefont{Blau et~al.}(2004{\natexlab{b}})\citenamefont{Blau,
  Frank, Loughlin, and Weiss}}]{Blau-Fran:2004}
\bibinfo{author}{\bibfnamefont{M.}~\bibnamefont{Blau}},
  \bibinfo{author}{\bibfnamefont{D.}~\bibnamefont{Frank}},
  \bibinfo{author}{\bibfnamefont{M.}~\bibnamefont{Loughlin}}, \bibnamefont{and}
  \bibinfo{author}{\bibfnamefont{S.}~\bibnamefont{Weiss}},
  \bibinfo{journal}{Paper in preparation}
  (\bibinfo{year}{2004}{\natexlab{b}}).

\bibitem[{\citenamefont{Cardoso
  et~al.}(2004{\natexlab{b}})\citenamefont{Cardoso, Natario, and
  Schiappa}}]{Cardoso-Nata:2004}
\bibinfo{author}{\bibfnamefont{V.}~\bibnamefont{Cardoso}},
  \bibinfo{author}{\bibfnamefont{J.}~\bibnamefont{Natario}}, \bibnamefont{and}
  \bibinfo{author}{\bibfnamefont{R.}~\bibnamefont{Schiappa}}
  (\bibinfo{year}{2004}{\natexlab{b}}), \eprint{hep-th/0403132}.

\bibitem[{\citenamefont{Natario and Schiappa}(2004)}]{Natario-Schia:2004}
\bibinfo{author}{\bibfnamefont{J.}~\bibnamefont{Natario}} \bibnamefont{and}
  \bibinfo{author}{\bibfnamefont{R.}~\bibnamefont{Schiappa}}
  (\bibinfo{year}{2004}), \eprint{hep-th/0411267}.

\bibitem[{\citenamefont{Witten}(1991)}]{Witten:1991}
\bibinfo{author}{\bibfnamefont{E.}~\bibnamefont{Witten}},
  \bibinfo{journal}{Phys. Rev.} \textbf{\bibinfo{volume}{D44}},
  \bibinfo{pages}{314} (\bibinfo{year}{1991}).

\bibitem[{\citenamefont{Mandal et~al.}(1991)\citenamefont{Mandal, Sengupta, and
  Wadia}}]{Mandal-Seng:1991}
\bibinfo{author}{\bibfnamefont{G.}~\bibnamefont{Mandal}},
  \bibinfo{author}{\bibfnamefont{A.~M.} \bibnamefont{Sengupta}},
  \bibnamefont{and} \bibinfo{author}{\bibfnamefont{S.~R.} \bibnamefont{Wadia}},
  \bibinfo{journal}{Mod. Phys. Lett.} \textbf{\bibinfo{volume}{A6}},
  \bibinfo{pages}{1685} (\bibinfo{year}{1991}).

\bibitem[{\citenamefont{Callan et~al.}(1992)\citenamefont{Callan, Giddings,
  Harvey, and Strominger}}]{Callan-Gidd:1992}
\bibinfo{author}{\bibfnamefont{C.~G.~J.} \bibnamefont{Callan}},
  \bibinfo{author}{\bibfnamefont{S.~B.} \bibnamefont{Giddings}},
  \bibinfo{author}{\bibfnamefont{J.~A.} \bibnamefont{Harvey}},
  \bibnamefont{and}
  \bibinfo{author}{\bibfnamefont{A.}~\bibnamefont{Strominger}},
  \bibinfo{journal}{Phys. Rev.} \textbf{\bibinfo{volume}{D45}},
  \bibinfo{pages}{1005} (\bibinfo{year}{1992}), \eprint{hep-th/9111056}.

\bibitem[{\citenamefont{Birmingham and Carlip}(2004)}]{Birmingham-Carl:2003}
\bibinfo{author}{\bibfnamefont{D.}~\bibnamefont{Birmingham}} \bibnamefont{and}
  \bibinfo{author}{\bibfnamefont{S.}~\bibnamefont{Carlip}},
  \bibinfo{journal}{Phys. Rev. Lett.} \textbf{\bibinfo{volume}{92}},
  \bibinfo{pages}{111302} (\bibinfo{year}{2004}), \eprint{hep-th/0311090}.

\bibitem[{\citenamefont{Ling and Zhang}(2003)}]{Ling-Zhan:2003}
\bibinfo{author}{\bibfnamefont{Y.}~\bibnamefont{Ling}} \bibnamefont{and}
  \bibinfo{author}{\bibfnamefont{H.-b.} \bibnamefont{Zhang}},
  \bibinfo{journal}{Phys. Rev.} \textbf{\bibinfo{volume}{D68}},
  \bibinfo{pages}{101501} (\bibinfo{year}{2003}), \eprint{gr-qc/0309018}.

\bibitem[{\citenamefont{Barvinsky et~al.}(2002)\citenamefont{Barvinsky, Das,
  and Kunstatter}}]{Barvinsky-Das:2002}
\bibinfo{author}{\bibfnamefont{A.}~\bibnamefont{Barvinsky}},
  \bibinfo{author}{\bibfnamefont{S.}~\bibnamefont{Das}}, \bibnamefont{and}
  \bibinfo{author}{\bibfnamefont{G.}~\bibnamefont{Kunstatter}},
  \bibinfo{journal}{Found. Phys.} \textbf{\bibinfo{volume}{32}},
  \bibinfo{pages}{1851} (\bibinfo{year}{2002}), \eprint{hep-th/0209039}.

\bibitem[{\citenamefont{Das et~al.}(2003)\citenamefont{Das, Ramadevi, Yajnik,
  and Sule}}]{Das-Rama:2002}
\bibinfo{author}{\bibfnamefont{S.}~\bibnamefont{Das}},
  \bibinfo{author}{\bibfnamefont{P.}~\bibnamefont{Ramadevi}},
  \bibinfo{author}{\bibfnamefont{U.~A.} \bibnamefont{Yajnik}},
  \bibnamefont{and} \bibinfo{author}{\bibfnamefont{A.}~\bibnamefont{Sule}},
  \bibinfo{journal}{Phys. Lett.} \textbf{\bibinfo{volume}{B565}},
  \bibinfo{pages}{201} (\bibinfo{year}{2003}), \eprint{hep-th/0207169}.

\bibitem[{\citenamefont{Dasgupta}(2003)}]{Dasgupta:2003}
\bibinfo{author}{\bibfnamefont{A.}~\bibnamefont{Dasgupta}}
  (\bibinfo{year}{2003}), \eprint{hep-th/0310069}.

\bibitem[{\citenamefont{Rovelli and Smolin}(1995)}]{Rovelli-Smol:1994}
\bibinfo{author}{\bibfnamefont{C.}~\bibnamefont{Rovelli}} \bibnamefont{and}
  \bibinfo{author}{\bibfnamefont{L.}~\bibnamefont{Smolin}},
  \bibinfo{journal}{Nucl. Phys.} \textbf{\bibinfo{volume}{B442}},
  \bibinfo{pages}{593} (\bibinfo{year}{1995}), \eprint{gr-qc/9411005}.

\bibitem[{\citenamefont{Ashtekar and Lewandowski}(1997)}]{Ashtekar-Lewa:1996}
\bibinfo{author}{\bibfnamefont{A.}~\bibnamefont{Ashtekar}} \bibnamefont{and}
  \bibinfo{author}{\bibfnamefont{J.}~\bibnamefont{Lewandowski}},
  \bibinfo{journal}{Class. Quant. Grav.} \textbf{\bibinfo{volume}{14}},
  \bibinfo{pages}{A55} (\bibinfo{year}{1997}), \eprint{gr-qc/9602046}.

\bibitem[{\citenamefont{Alekseev et~al.}(2003)\citenamefont{Alekseev,
  Polychronakos, and Smedback}}]{Alekseev-Poly:2000}
\bibinfo{author}{\bibfnamefont{A.}~\bibnamefont{Alekseev}},
  \bibinfo{author}{\bibfnamefont{A.~P.} \bibnamefont{Polychronakos}},
  \bibnamefont{and} \bibinfo{author}{\bibfnamefont{M.}~\bibnamefont{Smedback}},
  \bibinfo{journal}{Phys. Lett.} \textbf{\bibinfo{volume}{B574}},
  \bibinfo{pages}{296} (\bibinfo{year}{2003}), \eprint{hep-th/0004036}.

\end{thebibliography}

\end{document}